\documentclass[conference]{IEEEtran}
\usepackage[hyphens]{url}
\usepackage{hyperref}
\hypersetup{colorlinks,allcolors=blue}
\usepackage[backend=biber, style=ieee]{biblatex}
\usepackage{xurl}

\usepackage{amsmath,amssymb,amsfonts}
\usepackage{algorithmic}
\usepackage{graphicx}
\usepackage{textcomp}
\usepackage[table,xcdraw]{xcolor}
\usepackage{balance}
\usepackage{multirow}
\usepackage{amsmath}
\usepackage{array}
\usepackage{enumitem} 

\def\BibTeX{{\rm B\kern-.05em{\sc i\kern-.025em b}\kern-.08em
    T\kern-.1667em\lower.7ex\hbox{E}\kern-.125emX}}

\IEEEoverridecommandlockouts

\begin{document}

\title{Scalable and Ethical Insider Threat Detection through Data Synthesis and Analysis by LLMs}

\author{\IEEEauthorblockN{Haywood Gelman}
\IEEEauthorblockA{\textit{The Beacom College of Computer and Cyber Sciences} \\
\textit{Dakota State University}\\
Madison, SD, USA \\
haywood.gelman@trojans.dsu.edu}
\and
\IEEEauthorblockN{John D. Hastings}
\IEEEauthorblockA{\textit{The Beacom College of Computer and Cyber Sciences} \\
\textit{Dakota State University}\\
Madison, SD, USA \\
john.hastings@dsu.edu}
}

\maketitle
\begin{abstract}

Insider threats wield an outsized influence on organizations, disproportionate to their small numbers. This is due to the internal access insiders have to systems, information, and infrastructure. Signals for such risks may be found in anonymous submissions to public web-based job search site reviews. This research studies the potential for large language models (LLMs) to analyze and detect insider threat sentiment within job site reviews. Addressing ethical data collection concerns, this research utilizes synthetic data generation using LLMs alongside existing job review datasets. A comparative analysis of sentiment scores generated by LLMs is benchmarked against expert human scoring. Findings reveal that LLMs demonstrate alignment with human evaluations in most cases, thus effectively identifying nuanced indicators of threat sentiment. The performance is lower on human-generated data than synthetic data, suggesting areas for improvement in evaluating real-world data. Text diversity analysis found differences between human-generated and LLM-generated datasets, with synthetic data exhibiting somewhat lower diversity. Overall, the results demonstrate the applicability of LLMs to insider threat detection, and a scalable solution for insider sentiment testing by overcoming ethical and logistical barriers tied to data acquisition.
\end{abstract}


\section{Introduction}

Insider threats place organizations at great risk of attack due to internal employee access to systems, information, and infrastructure \cite{padayachee_joint_2021}. Research conducted on insider threats takes many paths to detection and mitigation through technological, psychological, behavioral, and educational tools \cite{greitzer_sofit_2018}. Despite available remedies, insider threat detection remains elusive due to the difficulty in perceiving technical and human indicators that signal the presence of an insider threat \cite{harms_exposing_2022,verizon_dbir_2024}.

One possible source of employee related sentiment is job search sites where current and past employees rate their employers. Usernames are anonymized, encouraging honest points of view in reviews \cite{zhang_best_2023}. Many reviews express negative sentiment toward current and former employers, which though hopefully honest, can cause damage to an employer's reputation \cite{hollig_online_2021,hollig_what_2024,zhou_financial_2019}. Considerable negative reviews on Glassdoor can also appear during times of a company’s monetary mismanagement \cite{hollig_online_2021,hollig_what_2024,zhou_financial_2019}, sentiment that pervades long after assuaging of monetary concerns \cite{zhou_financial_2019}.

The potential for anonymous employee reviews to reveal insider threat sentiment has seen minimal research. Reviews may express dissatisfaction or highlight organizational vulnerabilities, but some may also inadvertently or intentionally disclose confidential corporate data, proprietary secrets, or security risks~\cite{hollig_online_2021,zhang_best_2023}. This creates an opportunity to analyze such reviews for indicators of insider threats. However, ethical and legal concerns with web scraping and dataset validity~\cite{camargo-henriquez_web_2022,dai_bias_2024} pose challenges to leveraging these sources directly. In addition, through website restrictions, automation of non-authenticated access can produce only limited data \cite{karabarbounis_what_2018}. 

This research proposes that indicators of insider threats can be found in job reviews, including confidential and proprietary information that can harm an organization by intellect, finances, and reputation. Many existing insider threat detection methods use machine learning methods for sentiment analysis across various web platforms \cite{alsowail_empirical_2020}, but none, thus far, have used LLMs for this task. In addition, none of the work has looked at synthesizing data for insider threat research purposes. Inferred use can be drawn from \cite{shenoy_extended_2024} for cybersecurity detection and response, and the promise of refined sentiment analysis in \cite{mughal_comparative_2024}. To address these research challenges, this research utilizes LLMs to synthetically generate employer reviews through carefully crafted prompts \cite{shenoy_extended_2024} to minimize bias, thereby overcoming ethical and logistical barriers associated with sourcing data at scale through web scraping or other means. The following research questions guide the study:
\begin{enumerate}[label={\textbf{RQ\arabic*:}},left=1.0em]
    \item How can LLMs synthesize job site reviews that simulate human insider threat sentiment using carefully crafted prompts to minimize bias?
    \item How representative are the LLM synthesized review datasets compared to human data sets?
    \item How can LLMs be utilized to analyze synthetic and existing datasets for insider threat sentiment?
\end{enumerate}
This research is significant in its approach to uncovering insider threat sentiment in employer reviews, and expanding the scale of data available for insider threat research. This knowledge can assist in the formation of more effective corporate policy, to improve security of proprietary data, and increase public awareness of insider threats. By advancing the use of LLMs for security-focused sentiment analysis, this work contributes to a growing field of research and provides viable techniques for protecting organizations from internal threats.

\section{Methodology}

Claude Sonnet 3.5 \cite{claudeai_claude_2024} synthetically generated a job review dataset, and was compared to an aggregate pair of existing Glassdoor review datasets. Two LLMs, Sonnet 3.5 and GPT-4o \cite{chatgpt_chatgpt_2024}, were employed to analyze both datasets for insider threat sentiment. The methodology is composed of search criteria for related work, LLM prompts used for review generation and sentiment analysis, existing datasets employed, and methods used for sentiment analysis. The LLMs are used to analyze both synthetic and publicly available (yet limited) real-world data for insider threat sentiment, with results compared to human expert evaluations to assess validity.

\subsection{LLM Selection}
Research was performed on LLMs that met specific use criteria for review generation and sentiment analysis. Critical LLM criteria were natural language processing with native sentiment analysis capabilities, an API that could interface with Python to pass data from a comma separated value (CSV) file to the LLM, and the ability to write results back to a CSV file. For sentiment-related tasks, Sonnet 3.5 and GPT-4o are outstanding \cite{weitl-harms_using_2024}, while demonstrating that they can capably produce synthetic sentiment data \cite{hastings_utilizing_2024}. Both the tailored prompts described in this section, and the Python code for accessing the API, were adapted from \cite{hastings_utilizing_2024}. The code was adjusted for the current research by Claude and manually verified, while the prompts were adapted manually.

\subsection{Glassdoor Dataset}

Existing Glassdoor review datasets were chosen from Kaggle \cite{kaggle_dataset_glassdoor_nodate,gauthier_glassdoor_nodate} containing approximately 10.8 million records. Loosely defined insider threat related keywords including ``hate", ``toxic", ``caught", ``steal", ``corrupt", ``collu", ``stole", ``delet", ``pay", ``paid", and ``fraud" reduced the dataset to 1.8 million. The given approach involved selecting multi-match keywords (``delet" for ``delete", ``deletion", ``deleted", or ``deleting", ``collu" for ``colluding", ``colluded", or ``collusion") to minimize bias in keyword choices. For a population of 10.8 million records, a sample size of 385 (rounded up from 384.14) was determined based on \cite{cochran1977sampling}. The sorted order of the 1.8 million record set was randomized with the first 385 records chosen as the sample and used as the matching set. Matching is required in this situation because a control set is not possible. 

\subsection{Expert Scoring Criteria}

Human insider threat manual scoring of Glassdoor reviews was performed by one expert insider threat researcher as the gold standard, and used to create a suitable comparison to LLM-generated reviews. Criteria include scores 0.0 to 1.0 (0.0 as most negative, 1.0 as most positive) with general levels of critical, high, medium, low, and nominal, respectively. Critical level indicators (0.0-0.2) include revenge motivations, unambiguous intent to commit an insider threat act, and demonstration of knowledge to damage systems, information, infrastructure, organizational finances, or reputation. High level indicators (0.2-0.4) manifest as explicit objections to policy or managerial malfeasance, antipathy, and either a desire to seek employment elsewhere or a recognized bias after separation. Medium level indicators (0.4-0.6) include disturbance, minor policy disagreements, and specificity in criticism. Low level indicators (0.6-0.8) include productive disapproval, non-specific criticism, and statement of common concerns (pay complaints being the most common). Nominal level indicators (0.8-1.0) indicate neutral to positive employer attitudes with negligible insider threat risk. Crossover scores, where a score bordered between two scoring groups (0.2, 0.4, 0.6, and 0.8) were used to indicate a borderline threat determination between two groups.

Sentiment analysis prompts in this study intentionally avoid assigning a weight to scoring values so as not to bias the results. This was an important consideration in order to study the inherent understanding of LLM insider threat sentiment. The above level indicators determine that a neutral score indicates no insider threat risk in the 0.8-1.0 range. Single score results that follow will demonstrate the LLMs inherently understood this critical insider threat sentiment concept.

\subsection{Synthetic Reviews (RQ1)}

For synthetic review generation, Sonnet 3.5 was chosen based its performance in ad hoc testing. Using the API, 385 synthetic reviews were generated with the prompt in Table \ref{syn-prompt}. The prompt requests that Sonnet 3.5 generate a job review using a given sentiment score between 0.0 and 1.0, with 0.0 as most negative and 1.0 as most positive. The prompt was sent 35 times per each of the 11 scoring positions (i.e., 0.0, 0.1,..., 1.0) until 385 reviews were generated. To construct datasets with a similar configuration to the Glassdoor datasets, the prompt requests a randomly generated date, employee status, and job title, in addition to pros and cons that are common to Glassdoor reviews. The prompt was also given a word limit of 40 words per pro or con to minimize API cost, and asked to output the data in CSV-readable format for post-processing. 

\begin{table}[h!]
\caption{Insider Threat Synthesis Prompt}
\label{syn-prompt}
\centering
\begin{tabular}{|p{8.4cm}|}
\hline

``You will produce a hypothetical employer review from a hypothetical employee. I will give you a sentiment score between 0.0-1.0, where 0.0 is most negative and 1.0 is most positive. For each sentiment number, you should produce a review that is aligned with that sentiment. The review will consist of six components: Original sentiment, date of review (randomly generated with Gregorian date format M/D/YYYY between 1/15/2020 through 10/23/2024), employee status (current employee or former employee, randomly generated), job title (randomly generated, different from each other), pros, and cons. Pros and cons will be written in paragraph form and will be no more than 40 words in length each.   Output the data as a CSV with order data including orig\_sentiment, date\_of\_review, emp\_status, job\_title, pros, and cons. Use pipe as delimiter and include headers. Double quote all text fields.'' \\ \hline
\end{tabular}
\end{table}

\subsection{Text Diversity (RQ2)}

Text diversity is an analysis of the unique lexical characteristics of LLM-generated content \cite{hastings_utilizing_2024}, and can be scored through a variety of measurements including CR (compression ratio), CR-POS (part of speech compression ratio), and n-gram diversity score (NDS)~\cite{shaib_standardizing_2024}. Tests are designed to assess character, word combination, and sentence structure repetition within text~\cite{hastings_utilizing_2024}. This is of particular importance to LLM-generated text \cite{weitl-harms_using_2024} to determine how LLM data compares to human generated text when processing language \cite{shaib_standardizing_2024}. 

\subsection{Single Score Sentiment Analysis (RQ3)}

The Glassdoor and synthetic datasets were fed through the APIs for GPT-4o and Sonnet 3.5 to score insider threat sentiment in the reviews. Initially, the LLMs were asked to score the pros and cons fields separately. However, insider threat activity might appear in either the pros or cons fields, so generating one overall sentiment score by having the LLM look holistically at the full review  made more sense  than attempting to numerically combine pro and con scores. Therefore, the final prompt shown in Table \ref{sent-prompt} was used to produce single insider threat sentiment scores. In brief, this prompt requests that datasets be analyzed for insider threat sentiment contained in each review. A score range is provided to reflect positive or negative sentiment and a concise explanation for each sentiment score is requested. 

\begin{table}[h!]
\caption{Insider Threat Sentiment Analysis Prompt}
\label{sent-prompt}
\centering
\begin{tabular}{|p{8.4cm}|}
\hline
``You are an insider threat analyst. I will provide a job review consisting of pros and cons. Based on your implicit understanding of sentiment, analyze the job review for insider threat sentiment, and provide a sentiment score between 0.00-1.00 inclusive (to two decimal places) where is 0.00 is a completely negative sentiment and 1.00 is a completely positive sentiment. Include your confidence in the accuracy of the score to two decimal places. Additionally, provide a carefully crafted contextual explanation for the sentiment score that is related to the meaning of the text. Please provide your response in a text-based csv format on one line, with columns for the sentiment score, confidence, and explanation. Please do not provide any other response aside from the csv formatted data." \\ \hline
\end{tabular}
\end{table}

\subsubsection{Alignment Evaluation}
In assessing alignment between the gold standard scoring and LLM sentiment scoring of Glassdoor reviews, and between target sentiment scores and LLM sentiment scoring for synthetic reviews, well-established metrics for pairwise numerical data analysis were used: mean absolute difference (MAD) and mean squared difference (MSD) were employed\footnote{Mean absolute difference and mean squared difference are often referred to as mean absolute error and mean squared error, respectively, but ``difference" is mathematically proper in this context given the lack of a definitive ``correct" answer in the comparisons.}. These metrics provide a comprehensive analysis of outcomes in each approach.

\section{Results}

\subsection{Synthetic Reviews (RQ1)}

The approach described in the prior section was used by Sonnet 3.5 to synthesize reviews. An illustrative example containing a variety of interesting potential insider threats appears in Table \ref{job-review}. 

\begin{table}[h!]
\caption{Sample Job Review Synthesized by Claude Sonnet 3.5}
\label{job-review}
\centering
\begin{tabular}{|p{3.8cm}|p{4.2cm}|}
\hline
\textbf{Pros} & \textbf{Cons} \\ \hline
Building had a functional elevator most days. Water cooler was usually filled. & Data manipulated to hide problems. Forced to misrepresent findings to stakeholders. Severe burnout culture. \\ \hline
\end{tabular}
\end{table}

\subsection{Text Diversity (RQ2)}

Table \ref{tex-diversity} demonstrates text diversity results between Glassdoor job reviews and LLM-generated job reviews. The higher CR and CR-POS scores, along with lower NDS score for the Sonnet 3.5-generated job reviews indicates lower text diversity than Glassdoor, and hence less unique LLM-generated content \cite{weitl-harms_using_2024,hastings_utilizing_2024}. Text diversity is not the only measure of validity between the results that follow, especially in the minimally researched field of insider threat analysis using LLMs, with LLMs used to generate comparators. This concept will be addressed further in discussion and future research.

\begin{table}[h!]
\caption{Test diversity of Glassdoor and synthetically generated review datasets}
\label{tex-diversity}
    \centering
\begin{tabular}{ |p{1.5cm}||p{1.5cm}|p{1.5cm}|p{1.5cm}|  }
 \hline
  Dataset& CR &CR-POS&NDS\\
 \hline
 Sonnet 3.5   & 4.162    & 6.446 &   2.625 \\
 Glassdoor  & 2.647    & 4.779 &   2.857 \\
 \hline
\end{tabular}
\end{table}

\subsection{Glassdoor: Gold Standard vs. LLM Alignment (RQ3)}

The scores produced by GPT-4o and Sonnet 3.5 models on the Glassdoor reviews were checked for alignment by the expert gold standard scores, as demonstrated in Table \ref{expert-score-mad}. 

\begin{table}[h!]
\caption{Original Target and Evaluated Scores for the Gold Standard Dataset Scoring}
\label{expert-score-mad}
    \centering
\begin{tabular}{ |p{1.5cm}||p{1.5cm}|p{1.5cm}|p{1.5cm}|  }
 \hline
  Model& MAD &MSD&Max Diff\\
 \hline
 GPT-4o   & 0.256    & 0.096 &   0.75\\
 Sonnet 3.5  & 0.260    & 0.097 &   0.82\\
 \hline
\end{tabular}
\end{table}

This test demonstrated a maximum difference between expert scoring and LLM of 0.75 and 0.82 for GPT-4o and Sonnet 3.5, respectively. This large difference is accounted for by five reviews in particular, four by human error and one by a difference in human interpretation. Original Glassdoor single scores for these reviews were 0.9, 0.8, 0.85, and 0.8 respectively, incorrectly indicating neutral to positive insider threat sentiment. Two reviews were missing cons, which were initially recorded as a zero and corrected to 0.7 and 0.85, respectively. Blind corrected scoring determined values of 0.7, 0.35, 0.47, 0.42, 0.46, and 0.85. Table \ref{expert-rescore-mad} provides the updated results after blind rescoring.

\begin{table}[h!]
\caption{Corrected Target and Evaluated Scores for the Glassdoor Dataset}
\label{expert-rescore-mad}
    \centering
\begin{tabular}{ |p{1.5cm}||p{1.5cm}|p{1.5cm}|p{1.5cm}|  }
 \hline
  Model& MAD &MSD&Max Diff\\
 \hline
 GPT-4o   & 0.251    & 0.091 &   0.66\\
 Sonnet 3.5   & 0.254    & 0.091 &   0.73\\
 \hline
\end{tabular}
\end{table}

Analysis revealed that the human scorer missed six reviews, and made errors that led to the two largest MADs. These were corrected, but it doesn't rule out other human error. One concern remained where human scoring rated one review at 0.12 for a critical insider threat risk where GPT-4o and Sonnet 3.5 scored it at 0.75 and 0.85 for low to nominal risk. This review was initiated by a developer at a software company looking for a more challenging job, who stated, ``They pay very well at (redacted), but personally there's no amount of money that is ``enough" to me.'' This statement, coupled with the position and circumstances of the individual, make them a critical insider threat risk. 

\subsection{Synthetic Data: Target vs. LLM Alignment (RQ3)}

Table \ref{sent-scoring} shows the sentiment scoring of Sonnet 3.5 and GPT-4o for the review from Table \ref{job-review}. These evaluations are illustrative of the deep and nuanced understanding of insider threats seen throughout reviews generated and scored by LLMs. 

\begin{table}[h!]
\caption{Insider Threat Sentiment Scoring on Example Synthetic Review}
\label{sent-scoring}
\centering
\begin{tabular}{|p{0.7cm}|p{0.5cm}|p{6.3cm}|}
\hline
\textbf{LLM} &\textbf{Score} & \textbf{Explanation} \\ \hline
Sonnet 3.5 & 0.15 & This review exhibits strong negative insider threat indicators through mentions of data manipulation and forced misrepresentation to stakeholders, which suggests serious ethical compromises and potential security risks. The pros are minimal and purely infrastructural, while the cons reveal systemic integrity issues and employee duress that could motivate malicious insider actions. The burnout culture further amplifies risk by potentially pushing employees toward desperate measures. \\ \hline
GPT-4o & 0.10 & The review highlights minimal positive aspects that are trivial compared to the significant drawbacks. The positives involve basic conveniences that are expected in any work environment  while the negatives  such as data manipulation  forced misrepresentation  and burnout culture  are severe and critical issues that greatly overshadow any benefits. This creates a predominantly negative sentiment with high confidence in the analysis. \\ \hline
\end{tabular}
\end{table}

The insider threat sentiment scores produced by GPT-4o and Sonnet 3.5 on the synthetic reviews were checked for alignment with the target sentiment scores of the synthetic data. The purpose of this test is to determine comparative agreement between target sentiment and LLM scores for the 385 synthetic reviews. Where the Glassdoor tests used human scoring to provide a target sentiment for calculating absolute difference, these tests used the target sentiment (which was used to initially generate the synthetic data). The results, shown in Table \ref{synthetic-mad}, demonstrate greater congruity between the dataset and the LLMs than with human-scored reviews.

\begin{table}[h!]
\caption{Comparison of Target and Evaluated Scores for the Synthetic Dataset}
\label{synthetic-mad}
\centering
\begin{tabular}{ |p{1.5cm}||p{1.5cm}|p{1.5cm}|p{1.5cm}|  }
 \hline
  Model& MAD &MSD&Max Diff\\
 \hline
 GPT-4o   & 0.090    & 0.012 &   0.25\\
 Sonnet 3.5   & 0.089    & 0.011 &   0.25\\
 \hline
\end{tabular}
\end{table}

\section{Discussion}

\subsection{Sentiment Analysis}

Addressing RQ1, data analysis was expected to reveal a degree of bias, indicating limitations in current LLM development. Bias in synthetic reviews and sentiment analysis was minimized through careful prompt creation \cite{dai_bias_2024}. Human error is common in manual processing of large datasets. Aside from noted human errors, lessons learned can be applied to future work. Key findings determined that in many examples, expert analysis viewed indicators similarly to LLM sentiment analysis, and some examples differently. Similar scoring include an employer who demonstrated dishonest business dealings, discriminatory behavior, ethical and moral concerns exposed with specificity, a combative environment that creates disillusionment, threatening behavior from management, nepotism, and theft of code.  Differences in interpretation of indicators include the previously noted pay complaint (``there's no amount of money that is enough"), corporate revenge, evidence of dark trait characteristics of narcissism and psychopathy with disdain for people with education \cite{harms_exposing_2022}, and confrontational issues. One important area of commonality is that in agreement or disagreement, expert and LLM-generated reviews surfaced valid concerns of insider threats.

Indicators of insider threats in organizations are small, so a small degree of correlation was expected in the collected data between the two datasets, addressing RQ3. Results demonstrated indicators of insider threats proportional with levels of such threats observed within organizations through use of natural language processing inherent to LLMs. 

\subsection{Research Implications}

This research has applicability to significance and social construct by assisting organizations in gaining a better understanding of technical and reputational risks in job site reviews, as well as raising public awareness of insider threats.  The use of human and synthetic LLM-generated job reviews to test LLM-based sentiment analysis is effective based on an expected small degree of insider threat occurrence within organizations. A wide margin of disagreement was seen in only 1.3\% of reviews between human and LLM-generated sentiment analysis (5 out of 385), answering RQ3.

Addressing, RQ1 through RQ3, the research presented a repeatable methodology.  Generalizability lies in the global sources for the existing dataset, variables of interest, and the ability to reuse the given prompts so researchers may generate their own datasets. For global generalizability, prompts would be translated to a target language.

\section{Related Work}

\subsection{Trends}

Research by \cite{alzaabi_review_2024} in their review of insider threat research with natural language processing determine the viability of research using the CERT dataset~\cite{cmu_insider_2020}. Use of the CERT dataset is common among NLP insider threat researchers to establish a performance benchmark \cite{anul_haq_insider_2022,kumar_thee_2023,mittal_prediction_2023}, as well as researchers focusing on custom machine learning models \cite{he_insider_2021,apau_theoretical_2019}. Observed trends include manual \cite{weitl-harms_using_2024,soh_employee_2019} and automated labeling \cite{mittal_design_2023,moallem_understanding_2020,park_detecting_2018} of data fields for efficient data matching. Concerns are noted on accuracy of automated labeling. Technology trends lean on performance for insider threat models to improve detection \cite{yadav_sentiment_2022} and influence organizational policy \cite{osterritter_conversations_2021,randle_critical_2017,reegard_concept_2019}. Performance trending can be seen in LLM performance ranking by \cite{vardhni_performance_2024,metcalfe_enhancing_2024}.

\subsection{Challenges}

  \cite{weitl-harms_using_2024} researched measurement improvement of software approval ratings through NLP sentiment analysis by capturing statistical values from qualitative perceptions. Where \cite{weitl-harms_using_2024} used manual labels as the ``gold standard", \cite{park_detecting_2018} used automated labeling for insider threat machine learning analysis of secondary social media data, presenting research with label fidelity issues. \cite{camargo-henriquez_web_2022,mahto_dive_2016,yadav_sentiment_2022,nalawati_sentiment_2022,singh_enumerable_2022} applied web scraping techniques, NLP, and machine learning analysis to assist the research community, where their research is more appropriately applied to policy influence. \cite{soh_employee_2019} implemented an insider threat profiling mechanism based on sentiment analysis, and \cite{dai_bias_2024} addresses bias in data retrieval using LLMs.
    
\subsection{Gaps}

\cite{osterritter_conversations_2021} demonstrated policy influence, where \cite{cao_llm-assisted_2024} performed research on LLMs in public-sector decision making. Previously noted examples in the use of LLMs with sentiment analysis \cite{weitl-harms_using_2024,nalawati_sentiment_2022,singh_enumerable_2022,tan_unified_2019} along with research combining web scraping, LLM, and sentiment analysis was demonstrate the burgeoning state of the field \cite{camargo-henriquez_web_2022,yadav_sentiment_2022}. Use of LLM generated synthetic data is is observed by \cite{wu_exploring_2024} for improving job recommendations by reducing bias. \cite{skondras_generating_2023} applied synthetic data to improve job stratification. Similarly, \citeauthor{myronenko_improving_2024} utilized synthetic data to draw skill requirements for job roles from online job postings. 

By both generating synthetic datasets using LLMs and analyzing publicly available data, this study assesses the viability of synthetic data generation in this domain, and the comparative accuracy and validity of LLMs in identifying indicators of insider threats. Contributions of this research will address the use of synthetically generated job reviews to discover insider threat sentiment compared to a curated dataset. Policy influence to encourage organizational, informational, and reputational policy are its contributions.

\section{Conclusion}

Insider threats place organizations at grave risk of cyber, financial, and reputational risk. Job site reviews that contain indicators of threats to an organization's intellectual property, reputation, and cybersecurity provide an opportunity to gain anonymized insight into these threats. Insider threat research is a well-studied field, but little research exists on natural language processing of job site reviews that can cause degrees of harm to an organization by way of insider threats. Research utilized the the Sonnet 3.5 model to synthetically generate a dataset, as well as an existing dataset from Glassdoor, which were subsequently processed through Sonnet 3.5 and GPT-4o models for insider threat sentiment. Results were analyzed and compared using available data analysis tools to establish connections between collected data and insider threat sentiment. Results demonstrated a percentage of job reviews with insider threat sentiment proportional to the number of insider threats seen in organizations. The importance of this research is to create awareness for organizations as to their level of insider threat risk to influence policy in areas of employment, employee job satisfaction, data protection, protection of intellectual property, and protection of confidential information. Research provides a new insider threat dataset, openly available to insider threat researchers as a method to compare tool effectiveness. This research will also stimulate a dialog on the importance of insider threats, and to raise awareness with the public as to their criticality. The potential impact on policy will assist organizations in protecting information, improve employee job satisfaction, and protect organizations from financial and reputational loss.

\balance
\printbibliography

\end{document}